\documentstyle[11pt]{article}
\def\bdt{\dot \beta}
\def\adt{\dot \alpha}

\newfont{\bbbold}{msbm10 scaled \magstep1}

\def\bbC{\mbox{\bbbold C}}

\def\cG{{\cal G}}

\def\cN{{\cal N}}

\def\cV{{\cal V}}

\newfont{\goth}{eufm10 scaled \magstep1}

\def\gl{\mbox{\goth l}}

\def\gs{\mbox{\goth s}}

\def\a{\alpha}
\def\b{\beta}
\def\c{\gamma}\def\cdt{\dot\gamma}
\def\d{\delta}\def\D{\Delta}

\def\l{\lambda}
\def\m{\mu}
\def\n{\nu}
\def\p{\pi}
\def\P{\Pi}
\def\r{\rho}
\def\s{\sigma}\def\S{\Sigma}

\def\th{\theta}

\def\be{\begin{equation}}\def\ee{\end{equation}}
\def\bea{\begin{eqnarray}}\def\eea{\end{eqnarray}}
\def\ba{\begin{array}}\def\ea{\end{array}}

\def\del{\partial}

\def\xz{\times}

\def\del{\partial}

\let\la=\label

\def\bd{\begin{document}}
\def\ed{\end{document}}
\def\bea{\begin{eqnarray}}\def\barr{\begin{array}}\def\earr{\end{array}}
\def\eea{\end{eqnarray}}
\def\ft#1#2{{\textstyle{{\scriptstyle #1}\over {\scriptstyle #2}}}}
\def\fft#1#2{{#1 \over #2}}
\newcommand{\eq}[1]{(\ref{#1})}
\def\eqs#1#2{(\ref{#1}-\ref{#2})}
\def\det{{\rm det\,}}
\def\tr{{\rm tr}}
\newcommand{\ho}[1]{$\, ^{#1}$}
\newcommand{\hoch}[1]{$\, ^{#1}$}
\newcommand{\tamphys}{\it\small Center for Theoretical Physics,
Texas A\&M University, College Station, TX 77843, USA}
\newcommand{\newton}{\it\small Isaac Newton Institute for Mathematical
Sciences, Cambridge, UK}
\newcommand{\kings}{\it\small Department of Mathematics, King's College,
Strand, London WC2R 2LS, United Kingdom}

\newcommand{\auth}{\large P.S. Howe\hoch{a}, E. Sokatchev\hoch{b}
and P.C. West\hoch{a}}

\begin{document}
\input feynman

\hfill{KCL-TH-98-35}

\hfill{LAPTH-696-98}

\hfill{hep-th/9808162}

\vspace{20pt}

\begin{center}
{\Large{\bf 3-Point Functions in $\cN=4$ Yang-Mills}}
\vspace{30pt}

\auth

\vspace{15pt}

\begin{itemize}
\item [$^a$] \kings
\item [$^b$] {\it\small Laboratoire d'Annecy-le-Vieux de Physique
Th\'eorique\footnote{URA 1436 associ\'ee \`a l'Universit\'e de Savoie} LAPTH,
Chemin de Bellevue, B.P. 110, F-74941 Annecy-le-Vieux, France}
\end{itemize}

\vspace{60pt}

{\bf Abstract}

\end{center}

Three-point functions of analytic (chiral primary) operators in $\cN=4$
Yang-Mills theory in four dimensions are calculated using the harmonic
superspace formulation of this theory. In the case of the energy-momentum
tensor multiplet anomaly considerations determine the coefficient. Analyticity
in $\cN=2$ harmonic superspace is explicitly checked in a two-loop calculation.

{\vfill\leftline{}\vfill

\pagebreak
\setcounter{page}{1}

Although there was no known example of a four dimensional conformally invariant
quantum field theory in the 1960's and 1970's, the properties of such theories
were investigated. It  was realised that  conformal invariance could be used to
determine the  two- and three-point Green's functions up to constants in any
dimension  and the space-time dependence of many such correlators  were found
\cite{bucket}. With the discovery of supersymmetry, examples of conformally
invariant quantum field theory were found. The first such theory to be found
was the $\cN=4$ Yang-Mills theory \cite{finite}. Conformal invariance was very
successfully exploited \cite{bpz} in two dimensions to determine Green's
functions for higher-point functions in certain theories. However, these
developments in two dimensions  relied on the infinite nature of the two
dimensional conformal group and the existence of null vectors in certain
representations of this algebra. In four dimensions, the  conformal group is
only a finite dimensional group and  it appeared that, unlike in  two
dimensions, one would not be able to exploit conformal invariance to solve for
higher-point Green's functions. One indication to the contrary concerned the
Green's functions that involved $\cN=2$ chiral superfields of the same
chirality in the two dimensional $\cN=2$ minimal model series. Since these
correlators belong to a minimal conformal field theory, it was to be expected
that one could solve for these Green's functions explicitly, but in \cite{hw5}
it was  shown that one could do this using only the globally defined
superconformal group and chirality. As it is the globally defined part of the
two dimensional conformal group that generalises to higher dimensions, this
work lead to the hope \cite{hw},\cite{con} that the constrained nature of the
superfields that describe supersymmetric theories when combined with conformal
invariance might be sufficient, even in higher dimensions, to solve for more
than just the two- and three-point Green's functions. A related example of this
phenomenon is the simple relation between the anomalous weight of a chiral
superfield and its R weight in any superconformal theory, which in many cases
allows one to deduce the anomalous weight of the chiral superfield \cite{cw}.
\par
In fact all four dimensional supersymmetric theories of interest are described
by constrained superfields. The  Wess-Zumino model and the $\cN=1$ and $\cN=2$
Yang-Mills field strengths are described by chiral superfields. The remaining
theories of extended rigid supersymmetry are the $\cN=2$ matter and the $\cN=4$
Yang-Mills theory. The  $\cN=2$ matter is best described by a harmonic
superspace formulation \cite{har} in which it is represented by a single
component superfield $q^+$ which satisfies an analyticity condition. Here,
analytic means that $q^+$ is both Grassmann analytic, i.e. it depends on only
half of the fermionic coordinates in a similar manner to a chiral superfield,
and analytic on the internal (bosonic) space, a compact, complex manifold which
is used to extend standard ($\cN=2$) Minkowski superspace to the harmonic
superspace of interest. The $\cN=4$ Yang-Mills theory also has a succinct
description when formulated on an appropriate harmonic superspace
\cite{hh},\cite{hw}. Explicitly, the Yang-Mills field strength multiplet is
described by a single-component analytic superfield $W$ (taking its values in
the Lie algebra of $SU(N)$). Although, in the non-Abelian case, $W$ is
covariantly analytic (with respect to the gauge group),  the gauge invariant
operators, $A_q$, defined by

\be
A_q=\tr (W^q)
\la{1}
\ee

are analytic fields in the strict sense.

In references \cite{hw2,hw3,hw4} the constraints due to superconformal
invariance on four dimensional Green's functions involving chiral or harmonic
superfields were found. It was clear that these constraints were very strong
and it was suggested that they were sufficiently powerful to determine, up to
constants, a class of these Green's functions. These included  all the Green's
functions of operators composed of $\cN=2$ matter  of sufficiently low
dimension  in the $\cN=2$ supersymmetric theories and  operators composed of
gauge invariant  polynomials of the harmonic superfield $W$, also of
sufficiently low dimension, in the $\cN=4$ Yang-Mills theory. In the latter
case these are just Green's functions of the above operators $A_q=\tr (W^q)$
for sufficiently small $q$. The calculations required to establish this result
are complicated, but have been successfully completed for the four-point
Green's functions involving $\cN=2$ matter \cite{ehpsw}. This result encourages
us to believe that some four-point functions in the $\cN=4$ theory may be
amenable to a similar analysis. Some other four-point calculations from a
different viewpoint have appeared more recently \cite{lt,fmmr}.

Recently there has been considerable interest in the Maldacena conjecture which
relates string theory on $AdS$ backgrounds to conformal field theory on the
boundary \cite{mal}. In the most studied example it is conjectured that
classical IIB supergravity on $AdS_5\xz S^5$ is equivalent to the large $N$
limit of $\cN=4$ $SU(N)$ Yang-Mills theory on the boundary which in this case
is four-dimensional Minkowski spacetime. A key ingredient in this conjecture is
the fact that the symmetry groups of the supergravity background and the
conformal field theory are the same, namely $SU(2,2|4)$. Although not all
gauge-invariant operators in the $\cN =4$ theory are of the type given in
equation \eq{1} it turns out that it is precisely this set of operators that is
relevant to the Maldacena conjecture in its simplest form. The spectrum of IIB
supergravity on $AdS_5\xz S^5$ consists of the gauged $D=5, \cN=8$ supergravity
multiplet together with the massive Kaluza-Klein multiplets. These all fall
into short representations of $SU(2,2|4)$ with maximum spin 2 and are in
one-to-one correspondence with the superfields $A_q$ introduced above
\cite{af}.

An important example of this type of operator is the supercurrent
$T=A_2=\tr(W^2)$. This multiplet has 128 + 128 components and contains amongst
them the traceless, conserved energy-momentum tensor, four gamma-traceless,
conserved supersymmetry currents and fifteen conserved currents corresponding
to the internal $SU(4)$ symmetry of the theory.

In the present paper we shall focus on the two- and three-point functions of
the $\cN=4$ theory. On general grounds it is to be expected that one should be
able to solve for the two- and three-point functions of an arbitrary conformal
field theory in any dimension \cite{bucket}. However, the advantage of our
formalism is that it allows us to solve for the complete superfield correlation
functions in a very simple way, not least because the operators we are
interested in all have only one component. This means that the tensor
structures which arise in a component approach are dealt with automatically. In
addition, for those three-point functions which have non-zero leading terms in
a $\th$ expansion, it is easy to show that the solutions we obtain are unique.
In fact, the form of the two- and three-point function can be found as a
special case of the general formula given in ref. \cite{hw} for any Green's
function for which there are no corresponding superconformal invariants. The
procedure to determine certain of the higher-point Green's functions works in
essentially the same way, but the details are very much more complicated.

Analytic fields on harmonic superspace are most simply described in the setting
of complex spacetime \cite{hh,hw2}. In this setting they are holomorphic fields
on a complex superspace with $8$ even and $8$ odd coordinates. These
coordinates may be assembled into a supermatrix $X$ as follows:

\be
X=\left(\barr{cc}
x^{\a\adt} & \l^{\a a'}\\
\p^{a \adt} & y^{a a'}
\earr
\right)\ .
\la{11}
\ee

Here the indices $\a, \adt, a$ and $a'$ each take on
two values. The underlying body of this superspace is, locally, a
product of complex Minkowski space and an internal space which also has
four complex dimensions and which is coordinatised by the $y's$.
Locally, the internal
bosonic space is the same as complex Minkowski space but globally this is
not so, however, since one is usually interested in non-compact
Minkowski space, whereas the internal space is
always compact; in this instance it is the Grassmanian of 2-planes in
$\bbC^4$. From a computational point of view the $a$ and $a'$ indices
behave in exactly the same way as the two-component spacetime spinor
indices $\a$ and $\adt$.

The action of an infinitesimal superconformal transformation
on $X$ is given by

\be
\d X= B + A X + X D + XCX
\la{12}
\ee

where $\d g\in \gs\gl(4|4)$
(the complexified superconformal algebra) is given by

\be
\d g=\left(\barr{cc}
-A & B\\
-C & D\earr\right)
\la{13}\
\ee

and where the matrices $A, B, C$ and $D$ are now $(2|2)\xz(2|2)$
supermatrices. We note that there are 8 odd coordinates $\l^{\a a'}$
and $\p^{a\adt}$ whereas complexified $\cN=4$ super Minkowksi space
has 16. This means that fields defined on analytic superspace depend
on only half of the usual odd coordinates and are therefore to be
thought of as chiral in a generalized sense. The fields we shall
consider are also analytic in the internal $y$ coordinates; since the
internal part of the space is a compact complex manifold this means
that their dependence on these coordinates is severely restricted.

These fields are in fact holomorphic in all
the coordinates and are characterized by a positive integer $q$; under
superconformal transformations they transform as

\be
\d A_q = \cV A_q + q\D A_q
\la{14}
\ee

where $\cV$ is the vector field generating the transformation \eq{12} and where

\be
\D={\rm str}(A+XC)\ .
\la{15}
\ee

In this language the (free) field strength tensor $W$ is such a field with
charge $q=1$; in the non-Abelian case $W$ is not actually defined on this
superspace (rather it is covariantly analytic) but gauge-invariant operators of
the form

\be
A_q=\tr(W^q)
\la{16}
\ee

are analytic operators and transform as in equation \eq{14}. We observe that
these superfields define different short representations of $SL(4|4)$ depending
on the value of $q$ which must be integral. Assuming that quantum effects do
not disturb this representation structure $q$ must remain unchanged, and so
these fields will not have any anomalous dimensions because the dimension of
$A_q$, which is fixed by the above transformation law, is also given by $q$.
This is similar to the situation for chiral superfields where the dimensions
are determined by the $R$-charges \cite{cw}.

We now consider the two-point functions of such operators.
The basic building block is the two-point function for the free field
strength $W$; it is \cite{hw2}

\be
<W(1) W(2)>\  \propto\ g_{12}
\la{17}
\ee

where

\be
g_{12}=({\rm sdet} X_{12})^{-1}={\hat y_{12}^2\over x_{12}^2}
\la{18}\ .
\ee

Here $X_{12}=X_1-X_2$, etc, and the hatted $y$ variable is defined by

\be
\hat y_{12}= y_{12}-\p_{12}x_{12}^{-1}\l_{12}\ .
\la{19}
\ee

This variable is invariant under $S$-supersymmetry transformations.
The function $g_{12}$ can also be expressed in terms of a hatted $x$
variable which is $Q$-supersymmetric as

\be
g_{12}={y_{12}^2\over \hat x_{12}^2}
\la{20}
\ee

where

\be
\hat x_{12}=x_{12}-\l_{12}y_{12}^{-1}\p_{12}\ .
\la{21}
\ee

For the operators $A_q$ we find

\be
<A_{q_1}(1)A_{q_2}(2)>\ \propto\  \d_{q_1 q_2} (g_{12})^{q_1}
\la{22}\ .
\ee

In the case of the two-point function of two energy-momentum tensors $T=A_2$
the constant of proportionality can be determined by anomaly considerations as
we discussed below. It is essentially the central charge of the theory.

We now turn to the three-point functions. They are
expressions of the form

\be
<A_{q_1}(X_1) A_{q_2}(X_2) A_{q_3}(X_3)>\equiv \cG_{q_1 q_2 q_3}(X_1,X_2,X_3)\
.
\la{23}
\ee

On the assumption that analyticity is preserved in the quantum
theory\footnote{An explicit two-loop calculation confirming this assumption is
presented in the second half of the paper.}, the Ward Identity is

\be
\sum_{i=1}^3\left( \cV_i + q_i \D_i \right) \cG_{q_1 q_2 q_3}=0
\la{24}
\ee

where $\cV_i$ is the vector field generating a superconformal
transformation at the $i^{th}$ point. If the sum of the charges is
even, the solution to \eq{24}, up to an overall constant, is

\be
\cG_{q_1 q_2 q_3}=(g_{12})^{k_1} (g_{23})^{k_2} (g_{31})^{k_3}
\la{25}
\ee

where

\bea
k_1 &=& \ft12(q_1 + q_2 - q_3) \nonumber\\
k_2 &=& \ft12(q_2 + q_3 - q_1) \nonumber\\
k_2 &=& \ft12(q_3 + q_1 - q_2)\ .
\la{26}
\eea

The restriction to the sum of the $q$'s being even ensures that the $k$'s are
positive integers so that the solution is regular in the $y$'s. This must be
the case since each field can be expanded as a polynomial in $y$ with
coefficients which are fields defined on ordinary superspace. That \eq{25}
solves \eq{24} is easy to demonstrate. From \eq{17} we have

\be
(\cV_1 + \cV_2 + \D_1 + \D_2)g_{12}=0\ .
\la{27}
\ee

Using this and the values for the $k$'s given in \eq{26} we find
immediately that the Ward Identity \eq{24} is satisfied.

Furthermore, this solution is unique. Suppose there was another
solution, $\cG'$ say, then this could be written as $\cG \xz
{\cG'\over\cG}$ where $\cG$ is the above solution. But the ratio of
the two solutions would be an invariant under all superconformal
transformations of three points, and there are no such objects
\cite{hw3}.

If the sum of the charges is odd, it does not necessarily mean that the
corresponding correlation function should vanish because the charges are also
carried by the odd coordinates. Hence, in this case, the three-point functions
would be nilpotent. Furthermore, it is not so easy to establish uniqueness for
this type of correlation function since one cannot divide by nilpotent
quantities. Moreover, one might be able to multiply a given solution by a
function which is invariant up to terms that are annihilated by the nilpotent
leading term in the correlator.

We now compare our results for three-point functions with some of the partial
component results that have been given in the literature. Expressions for the
leading terms, i.e. the correlation functions of the Lorentz scalar fields
which form the leading components of each of the fields $A_q$, were given in
early work on conformal invariance \cite{bucket} and discussed in this context
in \cite{and}. In our formalism this Green's function is simply obtained from
the fully supersymmetric answer by dropping the hats on the $y$'s. If we denote
the leading scalars by $a_q(x,y)$ then we have the universal formula

\be
<a_{q_1} a_{q_2} a_{q_3}>=(g^0_{12})^{k_1}
(g^0_{23})^{k_2} (g^0_{31})^{k_3}
\la{28}
\ee

where

\be
g^0_{12}={y_{12}^2\over x_{12}^2}.
\la{29}
\ee

The structure of the $x$-factors in the denominator agrees with that expected
from reference \cite{bucket}.  The r\^{o}le of the $y$'s is to take care of the
group theory; when one expresses any field $a(x,y)$ explicitly in terms of
$y$'s and a scalar field with $SL(4)$ indices, one finds that the numerators of
three-point functions of fields of the latter type are given by combinations of
$SL(4)$ invariant tensors.

The next examples we shall consider concern Green's functions that contain  the
$SL(4)$ (complexification of $SU(4)$) current. The $SL(4)$ currents $J_{\m
i}{}^j$ appear at order $\l\p$ in the expansion of the superfield $T$.
Unfortunately, the spacetime derivatives of the leading scalars also appear at
this level, so that one has to separate out their contribution. Explicitly, one
has

\be
T\sim \l^{\a a'}\p^{a \adt}\left( \hat J_{\a\adt a a'} +\ft12
\del_{\a\adt}\del_{a a'}T_o\right)
\la{30}
\ee

where $T_o(x,y)$ is the leading component of $T$. The notation
here is that an $SL(4)$ superscript $i$ is replaced by a subscript $a$
and a superscript $a'$ while an $SL(4)$ subscript $i$ is replaced by a
superscript $a$ and a subscript $a'$. The $a$ and $a'$ indices are
raised and lowered using the epsilon tensor as usual. Thus we have

\be
J_i{}^j\rightarrow \left(\barr{cc}
J^a{}_b & J^{a b'} \\
J_{a' b} & J_{a'}{}^{b'} \earr\right)\ ,
\la{31}
\ee

with $J^a{}_a + J_{a'}{}^{a'}=0$. The field $\hat J^{a a'}$ is
then given by

\be
\hat J^{a a'}=J^{a a'} + y^{a b'}
J_{b'}{}^{a'} -J^a{}_b y^{b a'}- y^{a b'} J_{b' b} y^{b a'}\ .
\la{32}
\ee

We now extract from the $<TTT>$ the Greens function, the component that has
one $SL(4)$ current and two $T_0$ operators. This occurs in the coefficient of
$<TTT>$ that has one factor of $\l_1\p_1$, however we must take into account
the occurrence of the space-time derivatives of $T_0$ given into equation
\eq{30}.  One finds that

\be
<\hat J_{\a\adt a a'}(x_1,y_1)T_0(x_2,y_2)T_0(x_3,y_3)
\sim
y^2_{23}{(y^2_{31}(y_{12})_{a a'}
+y^2_{12}(y_{31})_{a a'})\over
x_{12}^2x_{23}^2x_{31}^2}({(x_{12})_{\a\adt}\over x_{12}^2}-
{(x_{31})_{\a\adt}\over x_{31}^2})
\ee

The $y$ factors just encode the correct $SL(4)$ group theory while the
space-time dependence agrees with that expected on grounds of just conformal
invariance for the amplitude which has one vector and two scalar operators
\cite{bucket} and discussed in this context in \cite{fand}.

Finally, we consider the component of $<TTT>$ which contains three $SL(4)$
currents. This will occur in the part of the $<TTT>$ that has  the factor
$\P_{i=1}^3(\l_i\p_i)$. However,  in order to extract it, we must as before
subtract out the contributions of the scalar operator $T_o$. It is
straightforward to do this, but somewhat lengthy, so we shall focus on showing
that the anomaly term in $<JJJ>$ is indeed present. One finds a term of the
form

\bea
<J_{\a\adt i_1}{}^{j_i}(x_1) J_{\b\bdt i_2}{}^{j_2}(x_2)
J_{\c\cdt i_3}{}^{j_3}(x_3)>&\sim&\Big\{\left(
\d_{i_1}{}^{j_2}\d_{i_2}{}^{j_3}\d_{i_3}{}^{j_1}+
\d_{i_2}{}^{j_1}\d_{i_3}{}^{j_2}\d{i_1}{}^{j_3}\right) \nonumber\\
&&-\ft12\left(
\d_{i_1}{}^{j_1}\d_{i_2}{}^{j_3}\d_{i_3}{}^{j_2}+
\d_{i_2}{}^{j_2}\d_{i_1}{}^{j_3}\d_{i_3}{}^{j_1}+
\d_{i_3}{}^{j_3}\d_{i_1}{}^{j_2}\d_{i_2}{}^{j_1}\right)\nonumber\\
&&+\ft14\d_{i_1}{}^{j_1}\d_{i_2}{}^{j_2}\d_{i_3}{}^{j_3}\Big\}
\xz f_{\a\adt\b\bdt\c\cdt}(x_1,x_2,x_3)
\la{33}
\eea

where the function $f$ is given by

\be
f_{\a\adt\b\bdt\c\cdt}(x_1,x_2,x_3)=
\left((x_{12})^4(x_{23})^4(x_{31})^4
\right)^{-1}\left((x_{12})_{\b\adt}(x_{23})_{\c\bdt}
(x_{31})_{\a\cdt}-(x_{12})_{\a\bdt}(x_{23})_{\b\cdt}(x_{31})_{\c\adt}\right)\
.
\la{34}
\ee

The combination of $SL(4)$ delta's in \eq{33} is simply
the $SL(4)$ $d$-tensor expressed in these indices while the function
$f$ is the anomaly triangle graph expression in two-component form.
Explicitly, the combination of $x$'s appearing in the numerator of $f$
is

\bea
(x_{12})_{\b\adt}(x_{23})_{\c\bdt}(x_{31})_{\a\cdt}-(x_{12})_{\a\bdt}
(x_{23})_{\b\cdt}(x_{31})_{\c\adt}&=&(\s_{\m})_{\a\adt}(\s_{\n})_{\b\bdt}
(\s_{\r})_{\c\cdt} \xz\nonumber\\ &&\tr\left(\c_5 \c_{\m}(\c\cdot x_{12})
\c_{\n}(\c\cdot x_{23})\c_{\r}(\c\cdot x_{31})\right)\nonumber\\ &&
\la{35}
\eea

Substituting \eq{34} into  \eq{33} and using \eq{35}
we find an expression that agrees with that expected from conformal
invariance alone \cite{bucket} and agrees with that given in
\cite{fand}. Although this procedure gives the functional forms for these correlation
functions it does not determine the overall coefficients in terms of the
coupling constant.

For supercurrent
correlators it turns out that the coefficients of two
and three-point functions are determined by their one-loop (free-field) values.
An argument demonstrating this non-renormalization theorem for two and three point correlators involving internal currents and the energy momentum tensor
was given in \cite{fand} using the results of \cite{afgj}.
We now give an alternative argument for the one-loop nature of the coefficients
of the two- and three-point functions of $T$. We are interested in $\cN=4$
Yang-Mills theory with the operators $T=\tr(W^2)$ as operator insertions. The
latter couples to the $\cN=4$ conformal supergravity multiplet. Hence if we
consider the quantum theory of $\cN=4$ Yang-Mills theory coupled to a classical
$\cN=4$ conformal supergravity background the correlation functions of $T$ will
automatically be included. Although the $\cN=4$ Yang-Mills theory is finite and
free of anomalies in flat space-time \cite{finite}, it is not in the presence
of $N=4$ conformal supergravity \cite{ft}.  However, it is known that in
perturbation theory this coupled theory is finite above one loop
\cite{hst},\cite{shelter}.  As a result, the  anomaly in the superconformal
symmetry has only a one-loop contribution. The three point function $<TTT>$
contains, for example, the three-point correlation function of the $SU(4)$
currents $J$. By taking the divergence of the latter we find that this
component of the three-point function is related to the axial anomaly. Since
the $<TTT>$ correlation function has only one unknown coefficient, this
coefficient must be the anomaly coefficient, up to a numerical factor. However,
as we have just discussed, the anomaly only has a one-loop contribution and as
a result the overall coefficient in $<TTT>$ is determined by this one-loop
contribution.

For correlators involving fields with higher values of $q$ it is not
immediately apparent that the above argument is applicable, but recent
calculations seem to suggest that non-renormalisation theorems may be true for
some three-point correlation functions of analytic operators \cite{df,and}. A
possible explanation for this is that the fields $A_q$ couple to the boundary
fields of the bulk AdS Kaluza-Klein supergravity multiplets. However, from a
ten-dimensional point of view all of these multiplets combine into the
ten-dimensional IIB supergravity multiplet so that one might suspect that this
common ten-dimensional origin could have some implications for the conformal
fields on the boundary.

In the rest of the paper we shall present an explicit calculation of the
two-loop contribution to the three-point function $\langle+2+3+3\rangle$ in an
${\cal N}=2$ theory consisting of complex (Fayet-Sohnius) hypermultiplets
coupled to Yang-Mills (as is well-known, if the matter is in the adjoint
representation, such a model describes ${\cal N}=4$ Yang-Mills in terms of
${\cal N}=2$ superfields). The main purpose of this example is to show that the
assumption of harmonic analyticity made earlier is indeed justified. We also
show that the two-loop contribution to this class of correlators actually
vanishes, in line with the results of ref. \cite{df}.

The ${\cal N}=2$ matter and Yang-Mills multiplets will be described in a way
which maintains the $SU(2)$ symmetry manifest \cite{har}. For this purpose we
introduce the Grassmann-analytic (G-analytic) harmonic superspace with
coordinates

\begin{equation}\label{6.1}
  x^\mu_A,\theta^{+\alpha}, \bar\theta^{+\dot\alpha}, u^\pm_i \ .
\end{equation}

Here $u^\pm_i$ are the harmonic variables parametrising the coset space
$SU(2)/U(1)\sim S^2$, i.e. one is to regard $u^\pm_i$ as the two columns of an
$SU(2)$ matrix; the index $i$ transforms under the (right) $SU(2)$ and $\pm$
are its harmonic (left) $U(1)$ projections. As a consequence, they have the
defining properties:

\begin{equation}\label{defu}
  u^-_i =(u^{+i})^*\;, \quad u^{+i}u^-_i = 1 \ .
\end{equation}

The Grassmann variables $\theta^{+\alpha},
\bar\theta^{+\dot\alpha}$ are $U(1)$ harmonic projections of the odd coordinates
of $N=2$ superspace,

\begin{equation}
  \theta^{+\alpha,\dot\alpha} = u^{+i}\theta_i^{\alpha,\dot\alpha} \ .
\end{equation}

The G-analytic space-time coordinate $x^{\alpha\dot\alpha}_A$ is obtained by
shifting $x^\mu$:

\begin{equation}
  x^\mu_A = x^\mu -2i \theta^{\alpha(i}\sigma^\mu_{\alpha\dot\alpha}
  \bar\theta^{\dot\alpha j)}u^+_iu^-_j \ .
\end{equation}

Under $Q$-supersymmetry it transforms into the $+$ projections $\theta^+$ and
not their complex conjugates $\theta^-$. This is the reason why the superspace
(\ref{6.1}) is called G-analytic. In what follows we shall always work in the
G-analytic superspace, therefore we shall drop the index $A$ of $x^\mu_A$.
Given two points in $x$ space, $x_{1,2}$, one can define the $Q$-supersymmetry
invariant difference

\begin{equation}\label{6.5}
  \hat x_{12} =x_{12} + {2i\over (12)} [(1^-2) \theta^+_1 \bar\theta^+_1 -
  (12^-) \theta^+_2 \bar\theta^+_2 + \theta^+_1 \bar\theta^+_2 + \theta^+_2
  \bar\theta^+_1] \ .
\end{equation}

Here $(12)$, $(1^-2)$, $(12^-)$ are short-hand notations for contractions of
harmonic variables:
  $$
  (12) \equiv u^{+i}_1 u^+_{2i} \;, \quad (1^-2) \equiv u^{-i}_1 u^+_{2i}
  \;, \quad (12^-)\equiv u^{+i}_1 u^-_{2i} \ .
  $$
In fact, $(12)$ and $\hat x_{12}$ are the $SU(2)$ covariant counterparts of the
variables $y_{12}$ and $\hat x_{12}$ from eq.(\ref{21}).

The matter and Yang-Mills multiplets are described by the analytic superfields
$q^+_r(x,\theta^+,\bar\theta^+,u)$ and $V^{++}_a(x,\theta^+,\bar\theta^+,u)$,
with $r$ and $a$ being indices of the matter and ajoint representations of the
gauge group, respectively. The details can be found in \cite{har}, here we only
give a brief summary of the Feynman rules \cite{frules}. The matter propagator
$\Pi$, the gluon propagator $P$ in the Fermi-Feynman gauge and the only vertex
relevant to our calculation are indicated below:

  \begin{center}
  \begin{picture}(0,3000)
  \drawline\fermion[\E\REG](-20000,0)[2000]
  \global\advance\pfrontx by -1200
  \global\advance\pbackx by 300
  \put(\pfrontx,\pfronty){\scriptsize 1r}
  \put(\pbackx,\pbacky){{\scriptsize 2s} \quad $\Pi_{12}$}

  \drawline\gluon[\E\REG](-7000,0)[3]
  \global\advance\pfrontx by -1200
  \global\advance\pbackx by 300
  \put(\pfrontx,\pfronty){\scriptsize 1a}
  \put(\pbackx,\pbacky){{\scriptsize 2b} \quad $P_{12}$}

  \drawline\fermion[\E\REG](5000,0)[4000]
  \global\advance\pmidy by -800
  \put(\pmidx,\pmidy){\scriptsize 1}
  \drawline\gluon[\N\REG](\pmidx,0)[2]
  \global\advance\fermionbackx by 1500
  \put(\fermionbackx,0){$g(t^a)^r_s\int du_1d^4x_1d^4\theta^+_1$}
  \end{picture}
  \end{center}
\vspace{3mm}
\centerline{Figure 1}
\vspace{5mm}

The expression of the matter propagator is

\begin{equation}\label{6.12}
  (\Pi_{12})^r_s = \langle \tilde q^{+r}(1)\vert q^+_s(2)\rangle =
  {(12)\over \hat x^2_{12}}\delta^r_s \ .
\end{equation}

Here we make use of the $Q$-invariant variable (\ref{6.5}). Eq. (\ref{6.12}) is
in fact the $SU(2)$ covariant counterpart of eq. (\ref{20}). The G-analyticity
of $\Pi_{12}$ is manifest, since only the $+$ projections of the Grassmann
variables appear. Note that the original form of the matter propagator given in
\cite{frules} is different, but the equivalent form (\ref{6.12}) is best suited
for our purposes in this paper. For the gluon propagator we shall use the
standard form from \cite{frules}:

\begin{equation}\label{glupr}
  (P_{12})_{ab} = \langle V^{++}_a(1)\vert V^{++}_b(2)\rangle =
  (D^+_1)^4\left({\delta^8(\theta_1-\theta_2)\over x^2_{12}} \right)
   \delta^{(-2,2)}(u_1,u_2)\delta_{ab} \ .
\end{equation}

Its G-analyticity with respect to the first argument is manifest, since it
contains the maximal number four of plus-projected spinor derivatives $D^+
= u^+_i D^i$ (just like the chiral matter propagators in ${\cal N}=1$
supersymmetry). G-analyticity with respect to the second argument is assured by
the presence of the Grassmann and harmonic delta functions which allow us to
transfer the spinor derivatives from point 1 to point 2.

Finally, the vertex describing the coupling of the gauge superfield to the
hypermultiplet is shown in Figure 1. It involves a G-analytic superspace
integral. Note that the harmonic integral must always be done after the
Grassmann one, since the analytic Grassmann measure $d^4\theta^+$ carries a
harmonic charge. The full Yang-Mills Feynman rules involve gluon vertices of
arbitrary order, as well as ghosts, but none of them show up at the two-loop
level.

The three-point function we want to compute involves gauge invariant composite
operators of harmonic charges $\langle+2+3+3\rangle$. The simpler case
$\langle+2+2+2\rangle$ turns out trivial, since the matter propagator
(\ref{6.12}) obey fermion type rules and a Furry-like theorem. The first
relevant graph is shown in Figure 2:
\vfill\eject

  \vspace{10 mm}
  \begin{center}
  \begin{picture}(0,0)
  \drawline\gluon[\E\REG](-2500,0)[5]
  \global\Xtwo=\pfrontx
  \global\Xthree=\pbackx
  \global\advance\Xtwo by -700
  \global\advance\Xthree by 300
  \put(\Xtwo,\pfronty){\scriptsize 5}
  \put(\Xthree,\pfronty){\scriptsize 4}
  \drawline\fermion[\NE\REG](\gluonfrontx,\gluonfronty)[3805]
  \global\advance\pbacky by 600
  \global\advance\pbackx by -200
  \put(\pbackx,\pbacky){\scriptsize 1}
  \drawline\fermion[\NW\REG](\gluonbackx,\gluonbacky)[3805]
  \drawline\fermion[\SW\REG](\gluonfrontx,\gluonfronty)[3500]
  \global\Xone=\pbackx
  \global\Yone=\pbacky
  \drawline\fermion[\SE\REG](\gluonbackx,\gluonbacky)[3500]
  \drawline\fermion[\E\REG](\Xone,\Yone)[10330]
  \put(\pmidx,\pmidy){\oval(10330,3000)[b]}
  \global\advance\pfrontx by -700
  \global\advance\pbackx by 300
  \put(\pfrontx,\pfronty){\scriptsize 3}
  \put(\pbackx,\pbacky){\scriptsize 2}
  \end{picture}
  \end{center}
  \vspace{15mm}
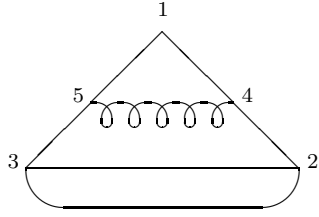
\centerline{Figure 2}
\vspace{5mm}

It should be remembered that this is a graph in $x$ space, therefore the true
loops are those involving the internal line 4-5, as opposed to the lines 2-3
which are just free propagators. Having this in mind and applying the Feynman
rules above, we find the corresponding expression (the gauge group indices and
factors are not shown):

\begin{equation}\label{ampl}
  I_1 = (\Pi_{23})^2 \int d^4x_{4,5}d^4u_{4,5}d^4\theta^+_{4,5}\;
   \Pi_{14} \Pi_{42}\Pi_{35}\Pi_{51}\;
  (D^+_4)^4\left({\delta^8(\theta_4-\theta_5)\over x^2_{45}} \right)
   \delta^{(-2,2)}(u_4,u_5) \ .
 \end{equation}

The first step in evaluating this graph consists in using the four spinor
derivatives $(D^+_4)^4$ from the gluon propagator to restore the full Grassmann
integral $\int d^4\theta^+_4 (D^+_4)^4 = \int d^8\theta_4$. This is made
possible by the explicit G-analyticity of the matter propagators $\Pi_{14}$ and
$\Pi_{42}$. Then the Grassmann $\delta^8(\theta_4-\theta_5)$ and harmonic
$\delta^{(-2,2)}(u_4,u_5)$ delta functions can be used to do the integrals
$\int du_4 d^8\theta_4$, thus identifying the Grassmann and harmonic points 4
and 5. In order to simplify the calculation, we shall evaluate the graph with
all the external Grassmann variables put to zero, $\theta_1 = \theta_2 =
\theta_3 = 0$. This corresponds to taking the lowest-order term in the $\theta$
expansion of the amplitude. This step allows us to easily deal with the hats
$\hat x$ in the matter propagators (see (\ref{6.5})). For the propagators
$\Pi_{23}$ the choice $\theta_{1,2,3}=0$ amounts to just removing the hat, but
for those involved in the vertex integrals, e.g. $\Pi_{14}$, there is still the
shift due to the integration variable $\theta_5$. Now, since the points 4 and 5
have been identified, the hats in all the propagators involve the same
Grassmann structure $\theta^+_5\bar\theta^+_5$ but different harmonic ones. All
this allows us to rewrite the amplitude (\ref{ampl}) as follows:

\begin{eqnarray}\label{amplatz}
  I_1(\theta_{1,2,3}=0) &=& {(23)^2\over x^4_{23}} \int du_5\;
(15)^2(25)(35)  \\
    && \int d^4\theta^+_5 \exp\left\{2i\theta^+_5\bar\theta^+_5 \cdot
  \left[{(15^-)\over (15)} \partial_1 + {(25^-)\over (25)} \partial_2 +
  {(35^-)\over (35)} \partial_3 \right] \right\}f(1,2,1,3)  \ .  \nonumber
\end{eqnarray}

Here $f(1,2,1,3)$ denotes the two-loop $x$-space integral

\begin{equation}
  f(1,2,1,3) = \int {d^4x_4 d^4x_5 \over
  x^2_{14} x^2_{24}x^2_{15}x^2_{35}x^2_{45}} \ .
\end{equation}

Using the translational invariance of $f(1,2,1,3)$ we can substitute
$\partial_1 f = -(\partial_2 + \partial_3)f$ in (\ref{amplatz}). Then we use
harmonic cyclic identities of the type $(25^-)(15) - (15^-)(25) = (12)$ (see
the defining property (\ref{defu})). Next we expand the exponential and do the
Grassmann integral, after which (\ref{amplatz}) is reduced to (up to an overall
factor)

\begin{equation}\label{ampint}
 I_1(0) = {(23)^2\over x^4_{23}} \int du_5 \left[
 {(12)^2(35)\over (25)}\partial_2\cdot\partial_2 +
  {(13)^2(25)\over (35)}\partial_3\cdot\partial_3 +
  (12)(13)\; 2\partial_2\cdot\partial_3 \right]f(1,2,1,3)  \ .
\end{equation}

The harmonic integral of the third term in (\ref{ampint}) is trivially done
($\int du_5\; 1 = 1$), and the first two ones are computed as follows (see
\cite{frules} for details):

$$
 \int du_5\;{(35)\over (25)} = \int du_5\;{\partial^{++}_5(35^-)\over (25)} =
 \int du_5\; (35^-)\delta^{(-1,1)}(u_2,u_5) = (32^-) \ .
$$

Further, the box operators in (\ref{ampint}) reduce the two-loop integral $f$
to a one-loop one, e.g.

$$
\partial_2\cdot\partial_2 \int {d^4x_4 d^4x_5
\over x^2_{14} x^2_{24}x^2_{15}x^2_{35}x^2_{45}} = {4\pi^2 i\over x^2_{12}} \int
{d^4x_5 \over x^2_{15} x^2_{25}x^2_{35}} \equiv {g(1,2,3)\over x^2_{12}} \ .
$$

The end result of all this is:

\begin{equation}\label{ampfin}
 I_1(0) = {(23)^2\over x^4_{23}} \left[
 (12)^2(32^-) {g(1,2,3)\over x^2_{12}}  +
  (13)^2(23^-){g(1,2,3)\over x^2_{13}}  +
  (12)(13)\; 2\partial_2\cdot\partial_3 f(1,2,1,3)\right]  \ .
\end{equation}

We immediately remark the presence of negative-charged harmonics in
(\ref{ampfin}) which means that the contribution of the graph in Figure 2 {\it
is not harmonic analytic}. However, the situation changes when we take into
account the two other graphs of similar type shown in Figure 3:

  \begin{center}
  \begin{picture}(0,5000)
  \drawline\gluon[\SW\REG](-8000,0)[3]
  \global\Xtwo=\pfrontx
  \global\Ythree=\pbacky
  \global\advance\Xtwo by 500
  \global\advance\Ythree by -750
  \put(\Xtwo,\pfronty){\scriptsize 4}
  \put(\pbackx,\Ythree){\scriptsize 5}
  \drawline\fermion[\SE\REG](\gluonfrontx,\gluonfronty)[5350]
  \drawline\fermion[\NW\REG](\gluonfrontx,\gluonfronty)[3805]
  \global\Xfour=\pbackx
  \global\Yfour=\pbacky
  \global\advance\Yfour by 600
  \global\advance\Xfour by -200
  \put(\Xfour,\Yfour){\scriptsize 1}
  \drawline\fermion[\SW\REG](\pbackx,\pbacky)[9155]
  \global\Xone=\pbackx
  \global\Yone=\pbacky
  \drawline\fermion[\E\REG](\Xone,\Yone)[12947]
  \put(\pmidx,\pmidy){\oval(12947,4000)[b]}
  \global\advance\pfrontx by -700
  \global\advance\pbackx by 300
  \put(\pfrontx,\pfronty){\scriptsize 3}
  \put(\pbackx,\pbacky){\scriptsize 2}

  \drawline\gluon[\SE\REG](8000,0)[3]
  \global\Xtwo=\pfrontx
  \global\Ythree=\pbacky
  \global\advance\Xtwo by -800
  \global\advance\Ythree by -750
  \put(\Xtwo,\pfronty){\scriptsize 4}
  \put(\pbackx,\Ythree){\scriptsize 5}
  \drawline\fermion[\SW\REG](\gluonfrontx,\gluonfronty)[5350]
  \drawline\fermion[\NE\REG](\gluonfrontx,\gluonfronty)[3805]
  \global\Xfour=\pbackx
  \global\Yfour=\pbacky
  \global\advance\Yfour by 600
  \global\advance\Xfour by -200
  \put(\Xfour,\Yfour){\scriptsize 1}
  \drawline\fermion[\SE\REG](\pbackx,\pbacky)[9155]
  \global\Xone=\pbackx
  \global\Yone=\pbacky
  \drawline\fermion[\W\REG](\Xone,\Yone)[12947]
  \put(\pmidx,\pmidy){\oval(12947,4000)[b]}
  \global\advance\pfrontx by 300
  \global\advance\pbackx by -700
  \put(\pfrontx,\pfronty){\scriptsize 2}
  \put(\pbackx,\pbacky){\scriptsize 3}
  \end{picture}
  \end{center}
  \vspace{23mm}
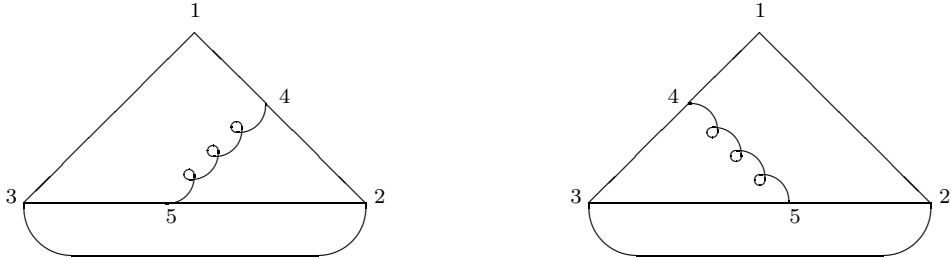
\centerline{Figure 3}
\vspace{5mm}

They can be computed in exactly the same way as the one in Figure 2. Putting
all three contributions together and using harmonic cyclic identities we find
that the non-analytic negative-charged harmonics disappear and we obtain the
{\it harmonic analytic} result:

\begin{equation}\label{endres}
  I(0) = I_1 + I_2 + I_3 = {(12)(13)(23)^2 \over x^2_{12} x^2_{13} x^4_{23}}
a(1,2,3)
\end{equation}

where

\begin{equation}\label{coefun}
 a(1,2,3) = x^2_{12}g(1,2,3)
   + x^2_{12} x^2_{13}\; 2\partial_2\cdot\partial_3 f(1,2,1,3)
   + \mbox{cycle}  \ .
\end{equation}

In (\ref{endres}) we observe a product of four matter propagators multiplied by
the coefficient function $a(1,2,3)$. The latter, according to the general
theory, must be a conformally invariant function of three points and hence can
only be a constant. In fact, it can be shown to vanish. A simple argument using
the Lorentz, translational and scaling properties of the integrals involved in
(\ref{coefun}) leads to the identity

$$
 2\partial_2\cdot\partial_3 f(1,2,1,3) =
 {x^2_{23} - x^2_{12} - x^2_{13}\over x^2_{12} x^2_{13}}\; g(1,2,3) \ .
$$

Substituting this in (\ref{coefun}) gives $a(1,2,3)=0$, so

\begin{equation}
  I(\theta_1=\theta_2=\theta_3=0)=0 \ .
\end{equation}

In other words, the lowest-order ($\theta=0$) term in the amplitude is zero,
and this can then be generalised to the entire amplitude.

The example of a three-point correlator presented above is not unique. It is
easy to construct three-point functions with higher $U(1)$ charges by simply
attaching more matter propagators to the external points. This does not affect
the loop structure of the graphs and thus leads to the same result as above.

In conclusion we should mention that there exist further two-loop graphs
involving the following insertions:

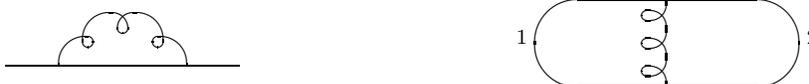
\vspace{5 mm}
  \begin{center}
  \begin{picture}(0,3000)
  \drawline\fermion[\E\REG](-15000,0)[2000]
  \drawloop\gluon[\NE 3](\pbackx,\pbacky)
  \drawline\fermion[\E\REG](\pbackx,\pbacky)[2000]
  \drawline\fermion[\W\REG](\pbackx,\pbacky)[7000]

  \drawline\gluon[\S\REG](10000,2500)[3]
  \put(\pmidx,\pmidy){\oval(10000,3300)[t]}
  \put(\pmidx,\pmidy){\oval(10000,3300)[b]}
  \global\advance\pmidx by -5700
  \put(\pmidx,\pmidy){\scriptsize 1}
  \global\advance\pmidx by 11000
  \put(\pmidx,\pmidy){\scriptsize 2}
  \end{picture}
  \end{center}
  \vspace{5 mm}

They need not be considered because both of them vanish (strictly speaking, the
second one is proportional to $\delta(x_{12})$, but in our analysis we always
keep the external points of an $n$-point function apart).

{\bf Acknowledgements}

The authors are grateful to Burkhard Eden, Eric D'Hoker, Dan Freedman and
Christian Schubert for discussions. This work was supported in part by the EU
network on Integrability, non-perturbative effects, and symmetry in quantum
field theory (FMRX-CT96-0012) and by the British-French scientific programme
Alliance (project 98074).
\vfill\eject

\end{document}